\documentclass[aip,apl,twocolumn,preprintnumbers,amsmath,amssymb]{revtex4}

\usepackage{graphicx}
\usepackage[colorlinks=true,citecolor=blue]{hyperref}

\begin{document}
    \title{Probing the ladder of dressed states and nonclassical light generation in quantum dot-cavity
    QED}
    \author{Arka Majumdar$^\dag$}
    \email{arkam@stanford.edu}
    \author{Michal Bajcsy$^\dag$}
    \email{bajcsy@stanford.edu}
    \author{Jelena Vu\v{c}kovi\'{c}}
    \affiliation{E.L.Ginzton Laboratory, Stanford University, Stanford, CA-$94305$\\
    $^\dag$ Equal Contributors}

\begin{abstract}
We investigate the photon induced tunneling phenomena in a
photonic crystal cavity containing a strongly coupled quantum dot
and describe how this tunneling can be used to generate photon
states consisting mainly of a particular Fock state. Additionally,
we study experimentally the photon-induced tunneling as a function
of excitation laser power and frequency and show the signature of
higher manifolds of the Jaynes-Cummings Hamiltonian in the
observed photon-statistics.

\end{abstract}
\maketitle

A single optical mode confined inside an optical cavity behaves
like a simple harmonic oscillator, where all the energy levels are
equally spaced. When this cavity mode is strongly coupled to a
two-level quantum emitter such as a quantum dot (QD), the energy
structure of the coupled system becomes anharmonic. This
anharmonic (Jaynes-Cummings) ladder has been recently probed in
atomic \cite{2009.PRL.Rempe.TwoPhoton} and super-conducting
\cite{Superconducting_dressed_states} cQED system. In addition,
nonclassical correlations between photons transmitted through the
cavity result from such anharmonicity, which in turn leads to
fundamental phenomena of photon blockade and photon induced
tunneling. These effects have been recently demonstrated in atomic
systems \cite{birnbaum_nature}, as well as solid-state
\cite{AF_natphys}. Moreover, photon blockade and photon-induced
tunneling can be used for applications beyond cQED, including
generation of single photons on demand \cite{AM_blockade_PRA} for
quantum information processing, high precision sensing and
metrology \cite{high-NOON}, as well as quantum simulation of
complex many-body systems \cite{ciuti_fermionized_photon}. In this
Letter, we explore the utility of the photon induced tunneling and
blockade for non-classical light generation and probing of higher
order dressed states in the solid state cQED system consisting of
a single quantum dot (QD) coupled to a photonic crystal cavity.
First, we provide numerical simulation data showing that photon
induced tunneling can be used to preferentially generate specific
multi-photon states. Following this, we present experimental data
demonstrating the transition from blockade to tunneling regime in
such a system and show the signature of higher order dressed
states observed in the measured photon statistics. The probing of
the ladder of dressed states by photon-correlation measurement has
previously been performed experimentally only in an atomic cavity
QED system \cite{2009.PRL.Rempe.TwoPhoton}, while in solid state
systems it has been studied theoretically
\cite{laussy_finley_dressed_state_2011} and signatures of higher
order dressed states were observed only using four wave mixing
\cite{second_order_langbein}.

\begin{figure}
\centering
\includegraphics[width=3.25in]{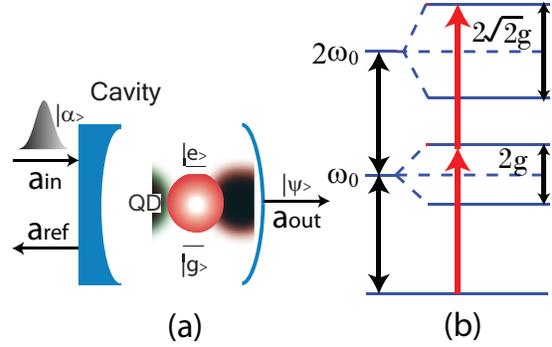}
\caption{(color online) (a) Schematic of the coupled QD-cavity
system driven by a Gaussian pulse (coherent state
$|\alpha\rangle$). The transmitted light through the cavity is
nonclassical($|\psi\rangle$) due to the nonlinearity provided by
the strongly coupled QD-cavity system. (b) The anharmonic
Jaynes-Cummings ladder structure.} \label{Figure1_setup}
\end{figure}

The dynamics of a coupled QD-cavity system, coherently driven by a
laser field (Fig. \ref{Figure1_setup} a), is well described by the
Jaynes-Cummings Hamiltonian of the form
\begin{equation}
\label{eqn:H} H=\Delta_a \sigma_+ \sigma_-+\Delta_ca^\dag
a+g(a^\dag\sigma_-+a\sigma_+)+\mathcal{E}(t)(a+a^\dag)
\end{equation}
\noindent which assumes the rotating wave approximation (RWA) and
a frame of reference rotating with the frequency of the laser
field $\omega_l$. Here, $\Delta_a=\omega_a-\omega_l$ and
$\Delta_c=\omega_c-\omega_l$ are respectively the detunings of the
laser from the QD resonant frequency $\omega_a$ and  from the
cavity resonance frequency $\omega_c$, $g$ is the coherent
coupling strength between the QD and the cavity mode,
$\mathcal{E}(t)=\sqrt{\kappa P(t)\over{2\hbar \omega_c}}$
\cite{supplementary} is the slowly varying envelope of the
coherent driving field with power $P(t)$ incident onto the cavity
with field decay rate $\kappa$, and $a$ is the annihilation
operator for the cavity mode. If the excited and ground states of
the QD are denoted by $|e\rangle$ and $|g\rangle$ then
$\sigma_-=|g\rangle\langle e|$ and $\sigma_+=|e\rangle\langle g|$.
Two main loss mechanisms in this system are the cavity field decay
rate $\kappa=\omega_{c}/2Q$ ($Q$ is the quality factor of the
resonator) and QD spontaneous emission rate $\gamma$. When the
coupling strength $g$ is greater than $\kappa\over2$ and $\gamma$,
the system is in the strong coupling regime
\cite{Yoshie04,Bloch_2005,Reithmaier_2004}. In this regime, energy
eigenstates are grouped in two-level manifolds with eigen-energies
given by $n\omega_c \pm g\sqrt{n}$ (for $\omega_a=\omega_c$),
where $n$ is the number of energy quanta in the coupled QD-cavity
system (Fig. \ref{Figure1_setup} b). The eigenstates can be
written as:
\begin{equation}
|n,\pm\rangle=\frac{|g,n\rangle \pm |e,n-1\rangle}{\sqrt{2}}
\end{equation}
Signatures of the photon blockade and tunneling can be detected
through photon-statistics measurements, such as the second-order
coherence function at time delay zero $g^{(2)}(0)=\frac{\langle
a^\dag a^\dag a a\rangle}{\langle a^\dag a\rangle ^2}$.
$g^{(2)}(0)$ is less (greater) than $1$ in photon blockade
(tunneling) regime, signifying presence of single (multiple)
photons in the light coming out of the coupled QD-cavity system.
$g^{(2)}(0)$ can be experimentally measured by Hanbury-Brown and
Twiss (HBT) setup, where coincidences between the photons are
detected \cite{AF_natphys}. Another important statistical quantity
is $n^{th}$ order differential correlation function
$C^{(n)}(0)=\langle a^{\dag n}a ^n\rangle-\langle a^\dag a\rangle
^n$, which provides a clearer measure of the probability to create
$n$ photons at once in the cavity \cite{2009.PRL.Rempe.TwoPhoton}.
Second order differential correlation function can also be
expressed as $C^{(2)}(0)=[g^{(2)}(0)-1]n_c^{2}$, where
$n_c=\langle a^\dag a\rangle$ is the average intra-cavity photon
number. Particularly for a weakly driven system ($n_c \ll 1$),
$C^{(2)}(0)$ becomes positive only when the probability of
two-photon state becomes significant compared to that of a
single-photon state, while a peak in $C^{(2)}(0)$ indicates
maximum probability of a two-photon state inside the cavity. As
the driving power increases, the peak in $C^{(2)}(0)$ shifts
towards detunings corresponding to maximum probability of exciting
higher manifolds, as described below and in the supplement.

Although the photon blockade and tunneling phenomena can be
observed under continuous wave (CW) excitation in a numerical
simulation \cite{supplementary}, for practical consideration it is
important to analyze the response of the cavity-QD system to a
pulsed driving field. In particular, the ability to measure the
photon statistics of the system's output during the actual
experiment is determined by the time resolution capabilities of
the single photon counters in the HBT setup, which in practice do
not allow for $g^{(2)}$ measurement of a CW-driven cavity-QD
system. A common way to overcome this limitation is to drive the
strongly coupled cavity-QD system with a train of weak, coherent
pulses of sufficiently narrow bandwidth \cite{AF_natphys}. We use
quantum trajectory method \cite{CarmichaelOpenSystems,
Quan_trajectory_dep} to analyze the pulsed driving of the coupled
QD-cavity system and find the resulting photon statistics
\cite{AM_blockade_PRA}. We also investigate the effect of pure QD
dephasing \cite{article:majumdar09} on the photon statistics and
observe that, even though the actual value of $g^{(2)}(0)$ is
affected due to dephasing, the qualitative nature of the
$g^{(2)}(0)$ spectrum remains same \cite{supplementary}.
As the non-classical state is collected from the cavity, only the
collapse operator corresponding to the cavity decay ($a$) is
monitored. A histogram is calculated based on the photon counts in
the cavity decay channel, and probability $P(n)$ for having
exactly $n$ photons in the system is found.
\begin{figure}
\centering
\includegraphics[width=3.25in]{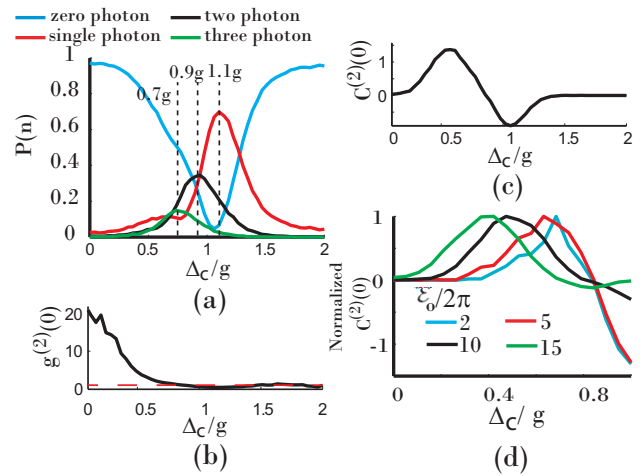}
\caption{(color online) Numerically calculated photon statistics
at the output of the QD-cavity system driven by Gaussian pulses
with duration $\tau_p \sim 24$ ps. The simulation parameters are
$g = 2\pi \times 40$ GHz, $\kappa =2\pi \times 4$ GHz,  and
$\mathcal{E}_o=2\pi \times 9$GHz; pure QD dephasing is neglected.
(a) $P(n)$, probability of generating an $n$ photon state at the
cQED system output as a function of laser-cavity detuning
$\Delta_c$. (b) Second order auto-correlation $g^{(2)}(0)$ as a
function of $\Delta_c$. The red dashed line shows the expected
$g^{(2)}(0)$ for a coherent state. (c) Second order differential
correlation $C^{(2)}(0)$ as a function of $\Delta_c$. (d)
$C^{(2)}(0)$ as a function of the laser-cavity detuning $\Delta_c$
for different values of the peak laser field $\mathcal{E}_o/2\pi$
(in units of GHz). We observe that the peak in the $C^{(2)}(0)$
occurs at $\Delta_c=0.7g$ for weaker excitation (where the second
order manifold is excited resonantly via two photons). However,
with increasing excitation power, the peak positions shifts
towards $\Delta_c=0$, due to excitation of higher manifolds.}
\label{fig2-pulsed_ideal}
\end{figure}
The driving term $\mathcal{E}(t)$ in the Hamiltonian described in
Eqn. \ref{eqn:H} is assumed to be of the form
$\mathcal{E}(t)=\mathcal{E}_o exp(-{t^2\over{2\tau_{p}^2}})$,
where $\mathcal{E}_o$ is the peak amplitude of the pulse. We set
$\tau_p=24.4$ ps (i.e., full width at half maximum - FWHM of $34$
ps), which satisfies the narrow-band condition  and corresponds to
our experimental parameters.

Figure \ref{fig2-pulsed_ideal} shows the behavior of the system
with better than current state of the art \cite{Arakawa_3d} but
achievable experimental parameters (assuming QD dipole moment of
$30$ Debye embedded in a linear three holes defect cavity with
mode volume $\sim 0.7(\lambda/n)^3$) resulting in $g = 2\pi\times
40$GHz and $\kappa = 2\pi\times 4$ GHz . These parameters can be
achieved by improving the alignment of the QD to the cavity field
and optimizing the photonic-crystal cavity fabrication process to
achieve higher quality factor. The results in
Fig.\ref{fig2-pulsed_ideal}a show that such a cavity-QD system can
be employed to deterministically generate selected Fock states of
high purity at the cavity output, where the particular Fock state
can be selected by adjusting the detuning of the drive laser from
the bare cavity resonance (no pure QD dephasing is included in the
simulation). The detuning values ($1.1g$, $0.9g$ and $0.7g$) are
different from what one intuitively expects from a lossless
strongly coupled QD-cavity system under CW driving ($g$,
$g/\sqrt{2}$ and $g/\sqrt{3}$, corresponding to the excitation of
first, second and third order manifold, respectively) because of
both the losses and the pulsed driving of the system
\cite{AF_natphys}. We note that, in presence of pure QD dephasing,
$P(n)$ for $n$ photon states decreases \cite{supplementary}. From
the probability distribution of the different Fock states we can
find the wave-function of the overall photon state
$|\psi\rangle=\sum\limits_{n}c_n|n\rangle$ with $P(n)=|c_n|^2$,
the second order coherence function
$g^{(2)}(0)=\frac{\langle\psi|a^{\dag}a^{\dag}aa|\psi\rangle}{\langle\psi|a^{\dag}a|\psi\rangle^2}=\frac{\sum\limits_n
n(n-1)P(n)}{\left(\sum\limits_n nP(n)\right)^2}$ and second order
differential correlation function
$C^{(2)}(0)=\langle\psi|a^{\dag}a^{\dag}aa|\psi\rangle-\langle\psi|a^{\dag}a|\psi\rangle^2=\sum\limits_n
n(n-1)P(n)-\left(\sum\limits_n nP(n)\right)^2$, which we can
measure experimentally. Figure \ref{fig2-pulsed_ideal}b shows
$g^{(2)}(0)$ as a function of $\Delta_c$, the laser detuning from
the empty cavity. The dashed line indicates the expected
$g^{(2)}(0)$ for a coherent state. Figure \ref{fig2-pulsed_ideal}c
shows $C^{(2)}(0)$ as a function of $\Delta_c$. $C^{(2)}(0)$
transitions from negative to positive value with decreased
detuning at $\Delta_c\sim0.9g$, thanks to the excitation of the
second manifold in the ladder when two photons are simultaneously
coupled into the cavity-QD system. Fig. \ref{fig2-pulsed_ideal}d
shows $C^{(2)}(0)$ as a function of $\Delta_c$ for different laser
excitation powers. We note that, the peak position changes
depending on the excitation laser power and at lower driving power
we observe the peak at $\Delta_c\sim0.7g$, where the second order
manifold is excited via two photons. With increasing power, the
higher (third and more) manifolds starts being populated, and the
peak in $C^{(2)}(0)$ subsequently shifts to smaller values of
detuning. In Fig.\ref{fig2-pulsed_ideal}c, the peak in
$C^{(2)}(0)$ is at a detuning of $\Delta_c\sim 0.5g$.
\begin{figure}
\centering
\includegraphics[width=3.5in]{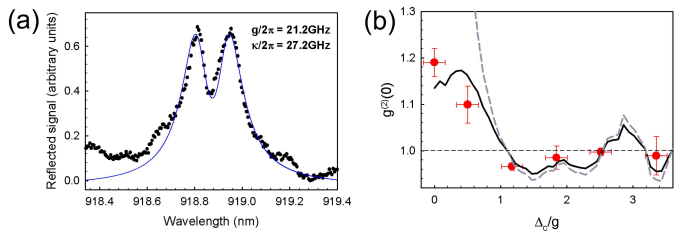}
\caption{(color online)(a) The transmission spectrum of a strongly
coupled QD-cavity system showing two polaritons. (b) Second order
coherence function at $t=0$, $g^{(2)}(0)$ as a function of the
laser detuning from the empty cavity frequency. The system is
excited with $\tau_p=24$ ps Gaussian pulses, with $80$ MHz
repetition frequency. The dashed grey (solid black) line results
from a numerical simulation based on the system's experimental
parameters and no (with) QD blinking. The average laser power for
the measurement is $P_{avg}=0.2$nW. For the simulations we use a
QD dephasing rate $\gamma_d/2\pi=1$ GHz. } \label{fig_exp1}
\end{figure}

\begin{figure}
\centering
\includegraphics[width=3in]{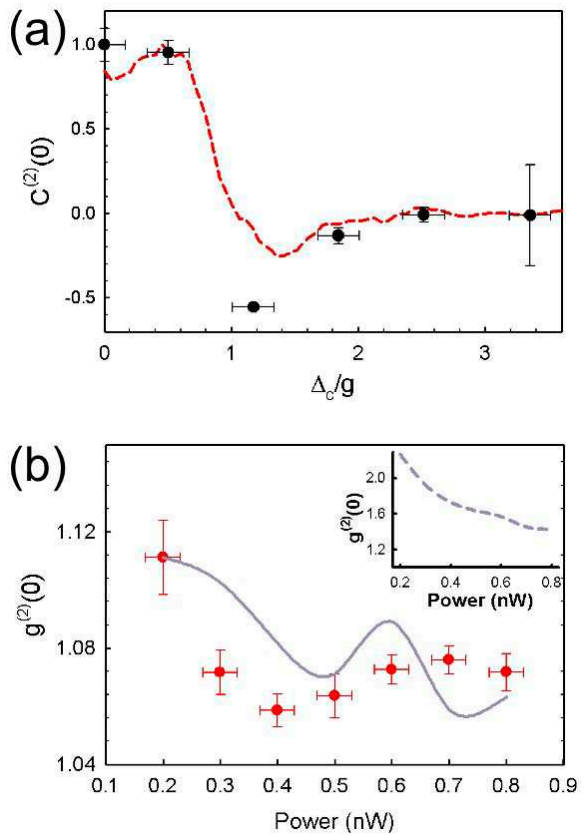}
\caption{(color online)(a) Normalized differential correlation
function $C^{(2)}(0)$ as a function of the laser detuning. The
dashed red line shows the result of a numerical simulation based
on the system's experimental parameters. (b) $g^{(2)}(0)$ in the
tunneling regime ($\Delta_c=0$) as a function of laser power
$P_{avg}$ measured in front of the objective lens. The solid line
shows the result of numerical simulation including the effects of
QD blinking, while the inset plots the numerically simulated
$g^{(2)}(0)$ in the absence of blinking. For the simulations we
use a QD dephasing rate $\gamma_d/2\pi=1$ GHz. } \label{fig_exp2}
\end{figure}

We confirm our theoretical predictions by performing experiment
with InAs QDs coupled to a linear three hole defect GaAs photonic
crystal cavity. Details of the fabrication and experimental setup
can be found in Ref. \cite{AF_natphys}. We measure laser
transmission through the system (using a cross-polarized
reflectivity setup \cite{AF_natphys}) and observe anti-crossing
between the QD and cavity (by changing temperature) signifying the
system is in the strong coupling regime. At resonance, the QD and
cavity mix to generate two polaritons, seen as two Lorentzian
peaks in Figure \ref{fig_exp1}a. By fitting the spectrum at
resonance we estimate the system parameters as $\kappa/2\pi=27$
GHz (corresponding to $Q\approx6,000$) and $g/2\pi=21$ GHz. To
drive the cavity-QD system, we use a mode-locked Ti-Sapphire
laser, that generates $3$ ps pulses at a repetition rate of
$f_{rep}=80$ MHz. These $3$ ps pulses are passed through a
monochromator to elongate the pulse in time domain, which results
in pulses with approximately Gaussian temporal profile of $34$ ps
FWHM, corresponding to $\tau_p=24.44$ ps (as in our theoretical
analysis). We determine the amplitude of the coherent driving
field using $ \mathcal{E}_o=\sqrt{{\eta
P_{avg}\over{4\pi^{1\over2}Q\tau_{p}f_{rep}\hbar}}}$
\cite{supplementary}, where $P_{avg}$ is the average optical power
of the pulse train measured before the objective lens and
$\eta\sim 0.03$ \cite{AF_natphys} is the coupling efficiency of
the incident light into the cavity including all the optics
losses. For our experimental parameters, $\mathcal{E}_o\approx
2\pi \sqrt{P_{avg}(nW)}\times9.3$GHz. The second order
auto-correlation $g^{(2)}(0)$ is measured as a function of
excitation laser frequency (Figure \ref{fig_exp1}b) to observe
transition from photon blockade to photon induced tunneling
regime. Typical histograms obtained for blockade and tunneling are
shown in the supplementary material. We estimate $g^{(2)}(0)$ as
the ratio of the coincidence counts at zero-time delay and
non-zero time delay.

Following this we calculate the second order differential
correlation function $C^{(2)}(0)$ for the coupled-QD cavity system
as a function of the laser-cavity detuning (Fig. \ref{fig_exp2}a).
We observe the transition of $C^{(2)}(0)$ from negative to
positive values and the onset of a peak at $\Delta_c\sim0.5g$
corresponding to the excitation of the higher order dressed
states. Simulations with our system parameters are shown by the
dashed line in Fig. \ref{fig_exp2} a). As explained before, the
peak in $C^{(2)}(0)$ does not correspond exactly to the resonant
excitation of the second order manifold via two photon process,
because of the additional excitation of the higher order
manifolds. All the measurements are performed at $14$ K. We note
that, in the simulation, $g^{(2)}(0)$ in the tunneling regime is
much larger than the experimentally measured value as a result of
QD blinking, which causes the experimentally collected data to be
a weighted average of transmission through an empty cavity and a
cavity with strongly coupled QD; in other words, blinking
effectively squashes the $g^{(2)}(0)$ curve towards $g^{(2)}(0)=1$
\cite{AF_natphys}. We model the blinking behavior of the QD by
assuming that during a unit time interval the QD is active for a
fraction $r$ and inactive for $(1-r)$ of the time. Thus the
$g^{(2)}(0)$ measured in the experiment will be a statistical
mixture of the coherent photon state (when QD is inactive, i.e.,
QD-cavity coupling $g=0$) and the correlated photons from the
coupled QD-cavity system \cite{supplementary}. We obtain good fit
to our experimental data with $r=0.65$. The vertical error bars in
all the figures are computed from the uncertainties in the fit of
the histogram data sets. The horizontal error bars are given by
the uncertainty in the measurement of the laser wavelength or the
laser power.

Finally, Fig. \ref{fig_exp2} b shows $g^{(2)}(0)$ as a function of
excitation laser power in the tunneling regime ($\Delta_c=0$).
This data is taken with the same cQED system on a different day,
when the cavity is red-shifted compared to the previous
measurements. For this particular experiment, the QD and the
cavity are resonant at $26$ K. This slightly higher temperature
might cause more QD dephasing, leading to a worse value of
$g^{(2)}(0)$ ($1.12$ as opposed to $1.2$ from the previous
measurement). Overall, $g^{(2)}(0)$ decreases with increasing
laser power as expected from the intuitive picture of QD
saturation at high driving power and the numerical simulation.
This clearly shows that the bunching observed in tunneling regime
is coming solely from the quantum mechanical nature of the
QD-cavity system, and not from any classical effect. We also
observe interesting oscillatory behavior in $g^{(2)}(0)$ as a
function of power. An oscillatory behavior is also observed in the
simulation that includes the effects of QD blinking. Without any
QD blinking, the simulation results show a mostly monotonically
decreasing $g^{(2)}(0)$ with increasing laser power (inset of Fig.
\ref{fig_exp2}b).

Finally, we would like to point out that these measurements have
been performed at the lowest $P_{avg}=0.2$nW that we can reliably
use, corresponding to $\mathcal{E}_o\approx 2\pi\times 4.2$GHz.
This roughly corresponds to the red plot in the theoretical Fig.
\ref{fig2-pulsed_ideal}d, where the peak in $C^{(2)}(0)$ is near
$\sim 0.5g$. This lower power limit is caused by the limited
mechanical stability of the cryostat and the low overall
efficiency with which we can couple the cavity photons into the
single photon counters in our HBT setup. The time needed to
perform the second-order coherence measurement increases
quadratically with decreasing $P_{avg}$ and for low powers the
cavity drifts out of focus before we can collect sufficient number
of coincidence counts.

In summary, we analyzed the photon induced tunneling regime in a
coupled QD-cavity system and proposed a scheme to use this system
for multi-photon state generation. In addition, we experimentally
characterized the second order coherence function $g^{(2)}(0)$ for
a coupled QD-cavity system as a function of laser-cavity detuning
and laser power. Using the experimental results of the photon
statistics measurement, we find the signature of the higher order
manifolds of the Jaynes-Cummings anharmonic ladder in the second
order differential correlation function $C^{(2)}(0)$.

The authors acknowledge financial support provided by DARPA, ONR,
NSF and the ARO; and useful discussion with Dr. Andrei Faraon. The
authors also acknowledge Dr. Pierre Petroff and Dr. Hyochul Kim
for providing the QD sample.

\bibliography{Tunneling_bibl_PR}
\end{document}